\documentclass[reprint, amssymb, amsmath, aps, superscriptaddress, showpacs, pra, longbibliography]{revtex4-1}
\usepackage{graphicx}
\usepackage{hyperref}
\begin{document}

\title{Antiferromagnetic ground state in the MnGa$_4$ intermetallic compound}
\author{V.Yu. \surname{Verchenko}}
\email{valeriy.verchenko@gmail.com}
\affiliation{Department of Chemistry, Lomonosov Moscow State University, 119991 Moscow, Russia}
\affiliation{National Institute of Chemical Physics and Biophysics, 12618 Tallinn, Estonia}

\author{A.A. \surname{Tsirlin}}
\email{altsirlin@gmail.com}
\affiliation{Experimental Physics VI, Center for Electronic Correlations and Magnetism, Institute of Physics, University of Augsburg, 86135 Augsburg, Germany}

\author{D. \surname{Kasinathan}}
\affiliation{Max Planck Institute for Chemical Physics of Solids, 01187 Dresden, Germany}

\author{S.V. \surname{Zhurenko}}
\affiliation{Department of Physics, Lomonosov Moscow State University, 119991 Moscow, Russia}
\affiliation{P.N. Lebedev Physics Institute, Moscow 119991,  Russia}

\author{A.A. \surname{Gippius}}
\affiliation{Department of Physics, Lomonosov Moscow State University, 119991 Moscow, Russia}
\affiliation{P.N. Lebedev Physics Institute, Moscow 119991,  Russia}

\author{A.V. \surname{Shevelkov}}
\affiliation{Department of Chemistry, Lomonosov Moscow State University, 119991 Moscow, Russia}

\begin{abstract}
Magnetism of the binary intermetallic compound MnGa$_4$ is re-investigated. Band-structure calculations predict  antiferromagnetic behavior in contrast to Pauli paramagnetism reported previously. Magnetic susceptibility measurements on single crystals indeed reveal an antiferromagnetic transition at $T_N=393$\,K. Neutron powder diffraction and $^{69,71}$Ga nuclear quadrupole resonance spectroscopy show collinear antiferromagnetic order with magnetic moments alligned along the [111] direction of the cubic unit cell. The magnetic moment of 0.80(3)\,$\mu_B$ at 1.5\,K extracted from the neutron data is in good agreement with the band-structure results.
\end{abstract}


\maketitle

\section{Introduction}

Despite their simple chemical compositions, binary compounds of $3d$ transition metals and $p$-elements show intricate physics. Experimental and theoretical reports on their magnetic and transport properties are often contradictory, as in the case of FeGa$_3$. The latter serves as a rare example of an intermetallic compound showing semiconducting rather than conventional metallic behavior. The formation of the narrow band gap at the Fermi level could be due to the hybridization of valence orbitals, or due to strong electronic correlations of Mott-Hubbard type.

In FeGa$_3$, electrical resistivity and Hall effect measurements reveal the semiconducting behavior with the band gap of 0.5\,eV~\cite{fega3-3}. Local-density (LDA) band-structure calculations arrive at the band gap of $0.3-0.5$\,eV too, suggesting a minor role of electronic correlations~\cite{fega3-1,fega3-2}. Moreover, the nonmagnetic behavior anticipated in this case is confirmed by the negative and almost temperature-independent magnetic susceptibility~\cite{fega3-3} and by the absence of the Zeeman splitting in room-temperature $^{57}$Fe M{\"o}ssbauer spectra~\cite{fega3-4}. This way, the nonmagnetic and semiconducting behavior predicted by LDA seemed to be confirmed experimentally. However, Yin \textit{et~al.}~\cite{fega3-5} conjectured that the Fe atoms may be magnetic within the semiconducting ground state. By introducing Coulomb correlations in a mean-field approach, they obtained Fe-Fe dimers with antiparallel spins, and the band gap that still conformed to the one observed experimentally. In agreement with these predictions, recent neutron powder diffraction experiments revealed a complex (and hitherto unresolved) antiferromagnetic structure of FeGa$_3$ that even persists above room temperature~\cite{fega3-6}.

Another example comes from the MnB$_4$ compound, where chains of Mn atoms feature alternating distances, such that Mn-Mn dimers are formed. Uncorrelated band-structure calculations interpret this dimerization as Peierls distortion accompanied by the formation of a pseudo gap at the Fermi energy~\cite{mnb4-1}. Therefore, MnB$_4$ should be semiconducting and nonmagnetic, but, similar to FeGa$_3$, electronic correlations can stabilize magnetism in this compound too. Signatures of cooperative magnetism were indeed observed~\cite{mnb4-3}, although not confirmed in independent studies~\cite{mnb4-2,mnb4-5}. A unified view on the ground state of MnB$_4$ may involve the competition between Peierls and Stoner mechanisms in avoiding the electronic instability caused by equidistant Mn atoms in the chains~\cite{mnb4-4}.

Given the strong sample dependence of thermodynamic and transport properties, a combination of bulk measurements and local probes is essential to determine the correct ground state of binary intermetallics. Here, we focus on another member of this family, MnGa$_4$, and challenge the existing non-magnetic scenario~\cite{tga4} that was solely based on thermodynamic measurements. Using neutron diffraction and nuclear quadrupole resonance (NQR) spectroscopy, we establish that MnGa$_4$ is in fact magnetically ordered, but, in contrast to FeGa$_3$ and MnB$_4$, correlation effects are not involved, as the magnetically ordered state can be obtained within LDA.

\begin{figure}
\includegraphics{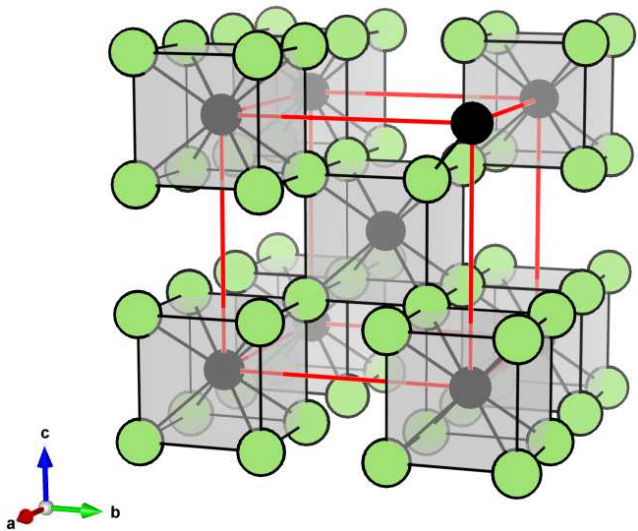}
\caption{\label{f1} MnGa$_{4}$ crystal structure with Mn atoms shown in black and Ga atoms shown in green. The unit cell is shown by the red lines. VESTA software~\cite{vesta} was used for crystal structure visualization.}
\end{figure}

MnGa$_4$ and its Cr-based analog CrGa$_4$ crystallize in the PtHg$_4$-type, which is a defect variant of the CsCl structure~\cite{tga4}. MnGa$_8$ cubes form the \textit{I}-centered cubic arrangement, in which 1/4 of Mn atoms and 3/4 of vacancies are fully ordered within the unit cell (Figure~\ref{f1}). In CrGa$_4$, the hybridization of Cr $3d$ and Ga $4s$ and $4p$ states opens a pseudogap at the Fermi energy. In the case of MnGa$_4$, the Fermi level is shifted to the conduction band, and the metallic ground state ensues~\cite{tga4}. Previous thermodynamic and transport measurements identified CrGa$_4$ as a diamagnetic bad metal, whereas Pauli paramagnetism and metallic conductivity were observed in MnGa$_4$~\cite{tga4}. In this study, we carry out a detailed investigation of MnGa$_4$ using high-quality single crystals, and unexpectedly find this compound to be antiferromagnetically ordered. We juxtapose our findings with the results of neutron diffraction, NQR measurements, and LDA calculations.

\section{Experimental details}

Crystals of MnGa$_4$ were grown from the high-temperature Ga flux. The mixture of Mn (5N, pieces, 99.999\%) and Ga (5N, pieces, 99.999\%) with the molar ratio $\nu\text{Mn}:\nu\text{Ga}=1:10$ was loaded into a quartz ampule, which was then evacuated to the residual pressure of $\approx1\times10^{-2}$\,torr and sealed. The ampule was annealed in a programmable furnace at 800\,$^{\circ}$C for 2 days, slowly cooled at the rate of 4\,$^{\circ}$C/h to 300\,$^{\circ}$C, and then cooled to room temperature in the shut-off regime. The excess of gallium metal was removed by centrifugation in an EBA 280 centrifuge (Hettich) at 40\,$^{\circ}$C. The obtained submillimeter-size crystals were further cleaned mechanically to remove traces of gallium. The bulk polycrystalline sample of MnGa$_4$ for neutron powder diffraction was prepared by annealling the stoichiometric mixture of Mn and Ga in an evacuated quartz ampule. The synthetic conditions were chosen on the basis of the reported phase diagram~\cite{mnga4}. The ampule was heated in a programmable furnace to 900\,$^{\circ}$C, annealed at this temperature for 4 days to ensure homogeneity of the mixture, cooled at the rate of 20\,$^{\circ}$C/h to 380\,$^{\circ}$C, and annealed at 380\,$^{\circ}$C for 10 days. Then, the sample was thoroughly ground and annealed at 380\,$^{\circ}$C for another 10 days.

Crystals of MnGa$_4$ were crushed by grinding. The resulting powder was mixed with Si (powder, 5N, 99.999\%) used as internal standard and analyzed on a Bruker D8 Advance powder diffractometer [Cu source, Ge (111) monochromator, $\lambda=1.540598$\, \r A]. The Rietveld refinement of the crystal structure was performed in the JANA2006 program~\cite{jana}.

Density-functional (DFT) band-structure calculations were performed using the FPLO code~\cite{fplo} (version 14.00-47). LDA version of the exchange-correlation potential~\cite{lda} was used in the scalar-relativistic regime. \textit{k}-space integration was performed by an improved tetrahedron method~\cite{int} on a grid of $16\times16\times16$ \textit{k}-points in both spin-restricted and spin-polarized calculations. Crystal orbital Hamilton population~\cite{cohp-1,cohp-2} (COHP) curves were calculated in the LOBSTER program (version 2.2.1)~\cite{lobster-1,lobster-2} using the band structure from VASP~\cite{vasp1,vasp2}.

Magnetization of MnGa$_4$ was measured on crystals, which were cleaned from traces of gallium metal mechanically rather than by the treatment with diluted HCl, thus preventing the formation of paramagnetic centers on the surface. For magnetization measurements, several crystals were glued together and measured as a polycrystalline sample. The data were collected using the Magnetic Properties Measurement System (MPMS, Quantum Design) at temperatures between 2\,K and 300\,K in magnetic fields of 0.1\,T, 0.5\,T and 5\,T. Measurements in the temperature range between 300\,K and 700\,K were performed using the VSM Oven Setup of a Physical Properties Measurement System (PPMS, Quantum Design) in 2\,T and 5\,T magnetic fields. Heat capacity was measured on several crystals glued together using a relaxation-type calorimeter (Heat Capacity option of PPMS) at temperatures between 1.8\,K and 50\,K in zero magnetic field. For resistivity measurements, a rectangular-shaped pellet with the dimensions of $0.8\times0.3\times0.2$\,cm$^3$ was pressed from powder at external pressure of 100\,bar. The relative density of 80\,\% was achieved. Cu wires with a diameter of 80\,$\mu$m were fixed on the pellet by hardening the silver-containing epoxy resin (Epotek H20E) at 100\,$^{\circ}$C. Resistance was measured by the standard four-probe technique in the temperature range 2--400\,K in zero magnetic field using the Resistivity option of PPMS.

Neutron powder diffraction (NPD) data were collected with the DMC diffractometer ($\lambda=4.5$\, \r A)  at 1.5\,K and 300\,K in the He cryostat, and with the HRPT diffractometer ($\lambda=1.886$\, \r A) at temperatures between 300\,K and 573\,K in the Nb high-vacuum oven at the Swiss spallation neutron source [SINQ, Paul Scherrer Institute (PSI), Villigen, Switzerland]. Rietveld refinements against the NPD data were performed with the JANA2006 program\cite{jana}.

The $^{69,71}$Ga nuclear quadrupole resonance (NQR) measurements were performed at 4.2\,K using a home-built phase-coherent pulsed NMR/NQR spectrometer. The $^{69,71}$Ga NQR spectra were measured using the frequency-step point-by-point spin-echo technique. At each frequency point, the area under the spin-echo profile was integrated in the time domain and averaged by a number of acquisitions.

\section{Results and Discussion}

\subsection{Synthesis and crystal structure}

\begin{figure}
\includegraphics{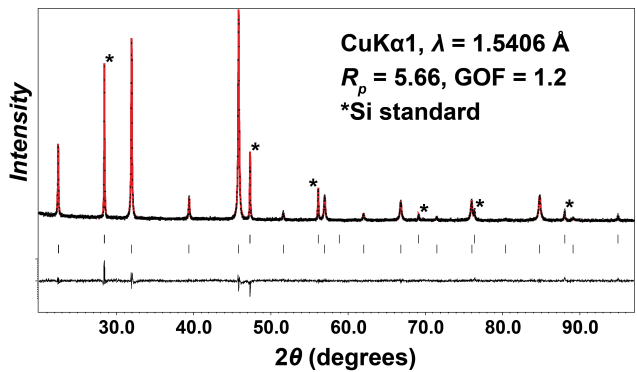}
\caption{\label{f2}Experimental (black points) and calculated (red line) PXRD patterns of MnGa$_4$. Peak positions are given by black ticks; the difference plot is shown as the black line in the bottom part. Peaks of the Si standard are marked with asterisks.}
\end{figure}

As gallium-rich compound, MnGa$_4$ can be grown from the high-temperature Ga flux. The synthesis yields small silvery-gray crystals of 0.1--1\,mm size. Powder x-ray diffraction (PXRD) confirms the formation of MnGa$_4$ as a single-phase product (Fig.~\ref{f2}). According to the Rietveld refinement against the PXRD data, MnGa$_4$ crystallizes in the PtHg$_4$ structure type, space group \textit{Im}-3\textit{m} (No.~229) with $a=5.59618(6)$\,\r A at room temperature. The crystal structure contains two crystallographic positions: Mn1 (0; 0; 0) and Ga1 (1/4; 1/4; 1/4). The refinement of site occupancies leads to the values of 0.997(7) and 1.003(7) for the Mn1 and Ga1 positions, respectively, thus confirming that the compound is stoichiometric.

\subsection{Thermodynamic and transport properties}

\begin{figure}
\includegraphics{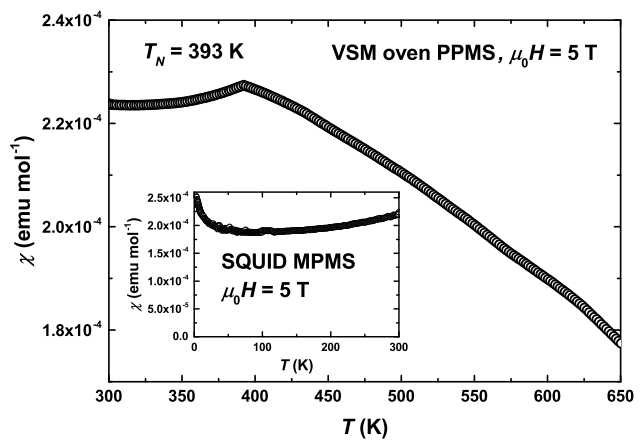}
\caption{\label{f3}Magnetic susceptibility $\chi(T)$ of MnGa$_4$ measured in the $\mu_{0}H=5$\,T magnetic field. The inset shows the data below room temperature.}
\end{figure}

The previous study reported MnGa$_4$ as Pauli paramagnetic metal with the temperature-independent magnetic susceptibility. No low-temperature anomalies were observed~\cite{tga4}. However, already the first magnetic susceptibility measurement in the temperature range of 2--300\,K (inset in Fig.~\ref{f3}) revealed that $\chi(T)$ is temperature-dependent. It shows an upturn at low temperatures as well as a slight increase around room temperature. The upturn below 75\,K is most likely due to paramagnetic impuritites, such as small amounts of defects in the real structure of MnGa$_4$. On the other hand, the observed increase of $\chi(T)$ above 100\,K is incompatible with the putative Pauli paramagnetism. Indeed, measurements above room temperature reveal an antiferromagnetic transition at $T_N=393$\,K. Above $T_N$, $\chi(T)$ decreases with increasing temperature, but does not follow the Curie-Weiss law up to at least 650\,K, whereas at 670\,K the decomposition of MnGa$_4$ occurs according to the reported phase diagram~\cite{mnga4}. Thus, magnetic susceptibility measurements suggest that MnGa$_4$ is antiferromagnetically ordered with the sizable Neel temperature of 393\,K. The high value of $T_N$ apparently concealed the antiferromagnetic nature of MnGa$_4$ in the previous study, where only measurements up to 300\,K were reported~\cite{tga4}.

\begin{figure}
\includegraphics{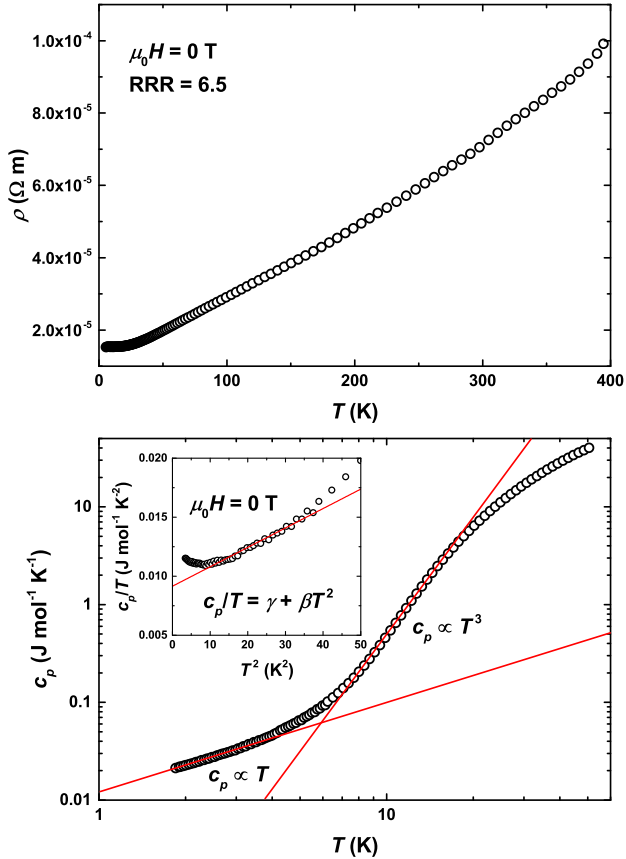}
\caption{\label{f4}(Top) Resistivity $\rho(T)$ of MnGa$_4$ measured in zero magnetic field. (Bottom) Heat capacity $c_p(T)$ of MnGa$_4$ in zero magnetic field. The inset shows the $c_p/T$ \textit{vs.} $T^2$ data at low temperatures.}
\end{figure}

The temperature-dependent resistivity and heat capacity of MnGa$_4$ are presented in Figure~\ref{f4}. Resistivity of MnGa$_4$ increases almost linearly with increasing temperature indicating good metallic conductivity in agreement with the previous report~\cite{tga4}. The observed small residual resistivity ratio (RRR) of 6.5 is probably due to the small relative density of the pressed pellet, which is 80\,\% of the theoretical value. 

The temperature-dependent heat capacity of MnGa$_4$ is reminiscent of metallic systems. Below 15\,K, it shows the $\propto T^3$ behavior due to lattice phonons, whereas below 4\,K the linear behavior driven by conduction electrons becomes prominent. The low-temperature part was fitted using the equation $c_p/T=\gamma+\beta T^2$, where $\gamma$ is the Sommerfield coefficient, and $\beta$ stands for the contribution of lattice phonons. The fit yields $\gamma=9.15(9)$\,mJ\,mol$^{-1}$\,K$^{-2}$ and $\beta=0.165(4)$\,mJ\,mol$^{-1}$\,K$^{-4}$, which is equivalent to the Debye temperature of $\Theta=389$\,K.

Altogether, we confirm the metallic behavior of MnGa$_4$, but additionally find that this compound should be antiferromagnetically ordered below $T_N=393$\,K. Other examples of antiferromagnetic intermetallic compounds include marcasite-type CrSb$_2$, a narrow-gap semiconductor with $E_g$ below 0.1\,eV~\cite{crsb2-2} and $T_N=273$\,K~\cite{crsb2-0}, as well as FeGa$_3$ with its putative incommensurate AFM order~\cite{fega3-6} and persistent AFM correlations in the metallic Co-doped regime~\cite{fega3-7}. The tendency to antiferromagnetism in these compounds is due to electronic correlations and can not be captured on the LDA level. In contrast, simple metals like elemental Cr and Mn order antiferromagnetically as a result of spin-density-wave instabilities. In the following, we set out to investigate whether the antiferromagnetic order in MnGa$_4$ is stable in LDA.

\subsection{Band structure and chemical bonding}

Band structure of MnGa$_4$ was calculated within the DFT framework. First, spin-unpolarized calculations were performed, and the dependence of the total energy on the unit cell parameter was calculated. The $E_{\rm tot}(a)$ plot displays a minimum at $a_0=5.481(2)$\, \r A. The calculated band structure at this equilibrium lattice parameter is in agreement with the previous study~\cite{tga4}, where strong hybridization between Mn and Ga valence orbitals was reported. The states between $-12$\,eV and 6\,eV are composed mainly of the Mn $3d$, Ga $4s$, and Ga $4p$ contributions. Mixing the Ga $4s$ and $4p$ orbitals leads to the bonding states at the energies of $-12<E-E_F<-4$\,eV (not shown), whereas high peaks of the density of states between $-3$\,eV and 1\,eV are due to the Mn 3\textit{d} -- Ga 4\textit{p} hybridization. 

\begin{figure}
\includegraphics{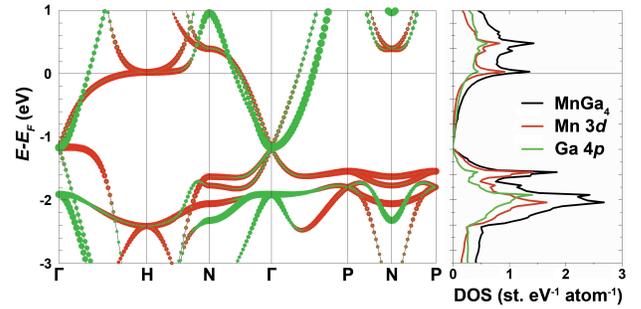}
\caption{\label{f5}(Left) LDA band structure of MnGa$_4$. (Right) Density of states plot calculated for MnGa$_4$. Contributions of the Mn $3d$ states and Ga $4p$ states are shown in red and green colors, respectively.}
\end{figure}

The states adjacent to the Fermi level are shown in Figure~\ref{f5}. Flat bands are seen at the Fermi energy and at relative energies between $-1.2$\,eV and $-2.4$\,eV where both Mn $3d$ and Ga $4p$ are present. At the same time, most of the parabolic bands have solely the Ga $4p$ character. Another feature of the Mn $3d$ -- Ga $4p$ hybridization is the formation of a direct pseudogap at the relative energy of $E-E_F=-1.2$\,eV. In the case of isomorphous CrGa$_4$, which has 18 valence electrons per formula unit (f.u.), the Fermi level is located directly in this pseudogap~\cite{tga4}. This situation -- the formation of a pseudogap in the band structure -- explains the stability of the PtHg$_4$-type intermetallic compounds that are formed when the number of valence electrons is 18 or 19 per f.u., according to the generalized $18-n$ rule\cite{rule}. In MnGa$_4$, which has 19 valence electrons per f.u., the Fermi level is shifted to the conduction band leading to the metallic behavior. The calculated density of states at the Fermi energy, $N(E_F)=0.84$\,st.\,eV$^{-1}$\,atom$^{-1}$, corresponds to the Sommerfield coefficient of the electronic specific heat $\gamma_{\rm bare}=9.9$\,mJ\,mol$^{-1}$\,K$^{-2}$. This value is in good agreement with the experimental value of $\gamma=9.15(9)$\,mJ\,mol$^{-1}$\,K$^{-2}$.

\begin{figure}
\includegraphics{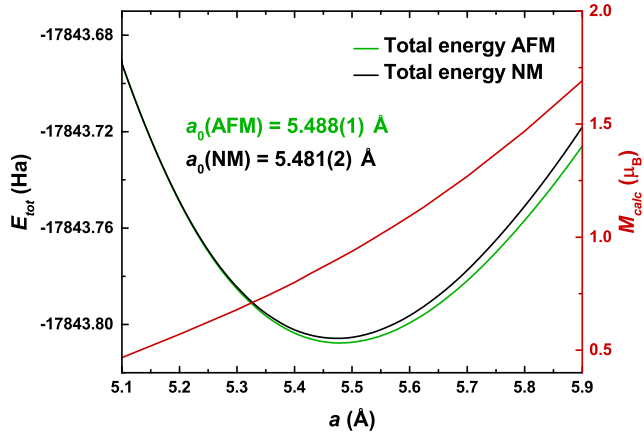}
\caption{\label{f6}(Left axis) Total energy of the antiferromagnetic (green line) and non-magnetic (black line) configurations of MnGa$_4$ plotted as $E_{\rm tot}(a).$ (Right axis) Calculated magnetic moment $M_{\rm calc}$ in the Mn1 position as a function of the unit cell parameter $a$.}
\end{figure}

Spin-polarized calculations reveal that the antiferromagnetic configuration, where magnetic moments point along the [111] direction, is energetically favorable in comparison with the ferromagnetic (FM) and non-magnetic (NM) configurations. The $E_{\rm tot}(a)$ curve calculated for the AFM case (Fig.~\ref{f6}) yields equilibrium values of $a_0=5.488(1)$\, \r A and $M_0=0.92$\,$\mu_B$. The calculated magnetic moment $M_{\rm calc}$ gradually decreases with decreasing the unit cell volume, which reflects the fact that antiferromagnetism of MnGa$_4$ may be suppressed under pressure. At ambient pressure, the predicted ground state of MnGa$_4$ is metallic and antiferromagnetic in agreement with the observed transport and thermodynamic properties.

Taking into account the itinerant nature of antiferromagnetism, chemical bonding analysis was performed using crystal orbital Hamilton population (COHP) method. A correlation between the bonding character of the states in the vicinity of the Fermi energy and the type of magnetic ordering has been proposed for itinerant systems~\cite{dronskowski}. When the states at the Fermi energy have purely antibonding character in spin-unpolarized calculations, ferromagnetic behavior ensues, leading to the rearrangement of the spin system in the way to leave these antibonding states empty. Alternatively, typical itinerant antiferromagnets show non-bonding states in the vicinity of the Fermi energy, and these states remain almost unchanged upon introducing spin polarization. Such an approach has been tested for the magnetic transition metals Cr, Mn, Fe, Co, and Ni as well as for multinary Fe-Mn rhodium borides~\cite{dronskowski}. 

\begin{figure}
\includegraphics{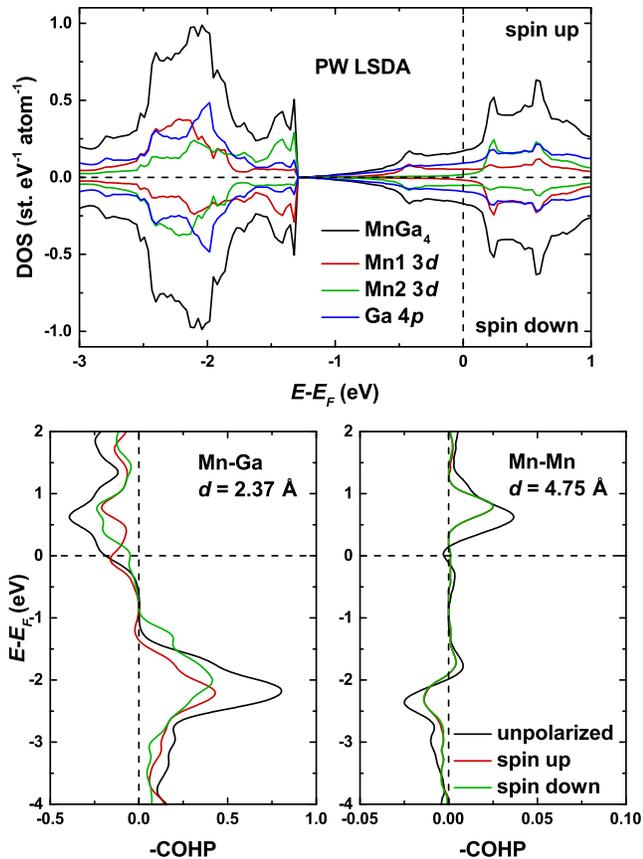}
\caption{\label{f7}(Top) Spin-polarized density of states plot obtained for the [111] direction of the magnetic moments. The Mn1 $3d$ contribution is shown in red color, the Mn2 $3d$ in green, and the Ga $4p$ in blue. (Bottom) Mn--Ga and Mn--Mn COHP curves calculated for MnGa$_4$. The spin-restricted COHP curves are shown in black color, spin-up in red, and spin-down in green.}
\end{figure}

In the MnGa$_4$ structure (Fig.~\ref{f1}), Mn atoms are connected with Ga atoms by short contacts with $d=2.37$\, \r A, while the shortest distances of $d=4.75$\, \r A between the Mn atoms are along the [111] direction. In the bottom panel of Figure~\ref{f7}, the corresponding COHP curves are shown. As expected, the short Mn-Ga contacts demonstrate negative values of $-$COHP at the Fermi energy and indicate the antibonding character of these states. This is rationalized by the fact that the Fermi level is located deep inside the conduction band formed as a result of the strong Mn-Ga hybridization. In MnGa$_4$, the valence and conduction bands are separated by a pseudogap located at $E-E_F=-1.2$\,eV. Accordingly, the states below this pseudogap show bonding character (positive values of $-$COHP), while the states above the pseudogap are antibonding.

The spin-resolved density of states plot calculated for MnGa$_4$ is shown in the top panel of Figure~\ref{f7}. The spin rearrangement occurs solely among the Mn $3d$ states leading to the formation of two magnetic subsystems located on the Mn1 and Mn2 atoms that compensate each other. As a result of spin polarization, the Mn-Ga states remain antibonding for the spin-up channel, while they tend to achieve the nonbonding character in the spin-down channel. At the same time, the Mn--Mn interactions show the nonbonding character in spin-unpolarized calculations, and introducing spin polarization has only minor effect, in agreement with the formalism proposed for itinerant systems~\cite{dronskowski}. The COHP curves reveal that the gain in the total energy within the antiferromagnetic state is triggered by the tendency of the Mn-Ga interactions in the spin-down channel to achieve the nonbonding character rather than remain antibonding. The interactions between the Mn atoms remain nonbonding upon introducing spin polarization and thus may not contribute to the total energy gain.

\subsection{Magnetic structure}

\begin{figure}
\includegraphics{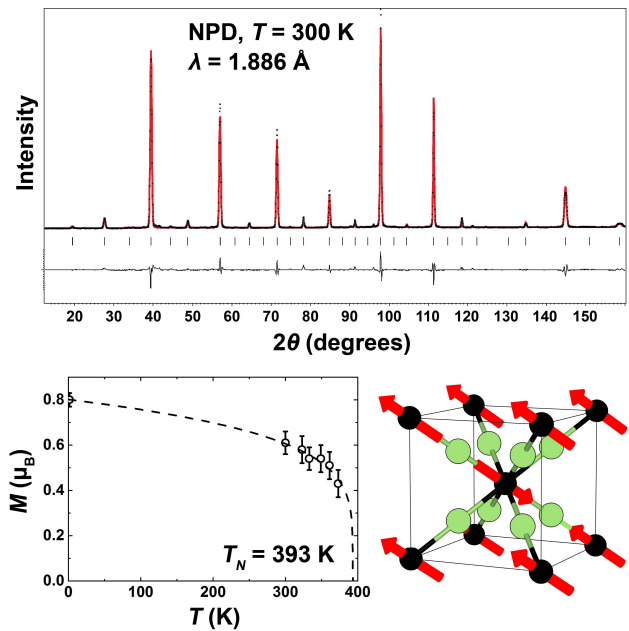}
\caption{\label{f8}(Top) Experimental (black points) and calculated (red line) NPD patterns of MnGa$_4$ at room temperature. Peak positions are given by black ticks, and the difference plot is shown as a black line in the bottom part. (Bottom, left panel) Magnetic moment $M$ on Mn atoms as a function of temperature. The dashed line is a guide to the eye. (Bottom, right panel) Magnetic structure of MnGa$_4$ as revealed by NPD.}
\end{figure}

Magnetic structure of MnGa$_4$ was investigated by neutron powder diffraction. The room-temperature NPD pattern is presented in Figure~\ref{f8}. The refinement shows that the magnetic and crystal lattices are commensurate, and $\vec{k}=0$. The best fit of the data was achieved for the model with antiparallel moments on the adjacent Mn atoms along the [111] direction. The magnetic moment of $M=0.61(5)$\,$\mu_B$ was obtained at $T=300$\,K. Further, it was found that the proposed magnetic model correctly describes the NPD data at all temperatures up to 393\,K. The extracted temperature dependence of the magnetic moment $M(T)$ is shown in the bottom left panel of Figure~\ref{f8}. At $T=1.5$\,K, the value of $M=0.80(3)$\,$\mu_B$ was obtained, which is only slightly smaller than the calculated (zero-temperature) value of $M_0=0.92$\,$\mu_B$. The remaining discrepancy may be due to spin fluctuations that are missing in LDA. With increasing temperature, $M$ decreases, and eventually the long-range AFM ordering dissappears at $T_N$.

\begin{figure}
\includegraphics{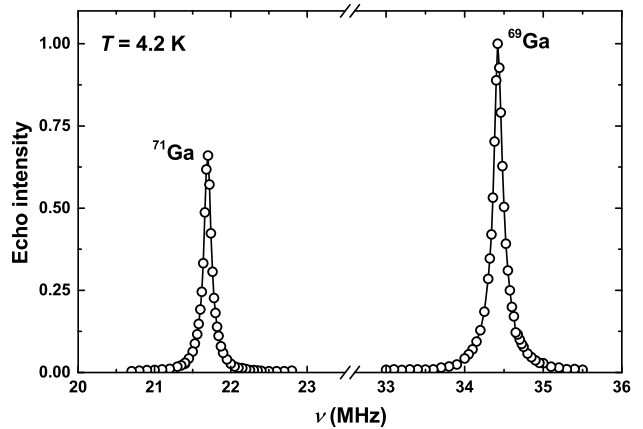}
\caption{\label{f9}$^{69,71}$Ga NQR spectrum of MnGa$_4$ measured at $T=4.2$\,K.}
\end{figure}

Magnetic structure of MnGa$_4$ was confirmed by $^{69,71}$Ga nuclear quadrupole resonance spectroscopy. The $^{69,71}$Ga NQR spectrum measured at 4.2\,K (Figure~\ref{f9}) shows two sharp signals centered at $\nu_1=21.7$\,MHz and $\nu_2=34.4$\,MHz. These two signals can be assigned to the $^{71}$Ga and $^{69}$Ga isotopes of Ga atoms that occupy one crystallographic position. Indeed, the intensity ratio of $I_1/I_2=0.66$ is in good agreement with the natural abundance of the isotopes, 60.11\,\% of $^{69}$Ga and 39.89\,\% of $^{71}$Ga. Also, the frequency ratio is equal to the ratio of the nuclear quadrupole moments, $\frac{\nu_1}{\nu_2}=\frac{eQ(^{71}\text{Ga})}{eQ(^{69}\text{Ga})}$, where $eQ(^{69}\text{Ga})=165.0(8)$\,mb and $eQ(^{71}\text{Ga})=104.0(8)$\,mb~\cite{nqr-ga}. The signals are sharp and feature the Lorentzian shape. This indicates a high degree of order in the Ga position, in particular, the absence of uncompensated local magnetic fields on the Ga nuclei. Therefore, $^{69,71}$Ga NQR spectroscopy confirms the collinear nature of the magnetic structure and corroborates the results of NPD.

\section{Conclusions}

MnGa$_4$ crystallizes in the PtHg$_4$ structure and shows no deviation from the stoichiometric composition. While resistivity measurements reveal metallic behavior, magnetic susceptibility indicates an AFM transition with the high Neel temperature of $T_N=393$\,K. The collinear $\vec{k}=0$ magnetic structure with magnetic moments directed along [111] is revealed by neutron diffraction. $^{69,71}$Ga NQR spectroscopy confirms the collinear nature of the magnetic order and the absence of uncompensated magnetic fields at the Ga site. This magnetic structure is reproduced by LDA, whereas the calculated zero-temperature magnetic moment $M_0=0.92$\,$\mu_B$ is only slightly higher than the experimental moment of $0.80(3)$\,$\mu_B$ at 1.5\,K. The COHP analysis reveals that the Mn-Ga interactions are bonding in the valence band and antibonding in the conduction band. The gain in the total energy upon spin polarization can be attributed to the fact that the Mn-Ga interactions achieve the nonbonding character in one of the spin channels. Altogether, we establish MnGa$_4$ as model itinerant antiferromagnet with the simple collinear magnetic structure that is well described within the single-electron approximation of LDA.

\begin{acknowledgements}
The authors thank Dr. Sergey Kazakov for his help with PXRD experiments and Christoph Geibel for suggesting the possibility of antiferromagnetic order in MnGa$_4$. We acknowledge the use of the Bruker D8 Advance X-ray diffractometer purchased under the Lomonosov MSU program of development. The work is supported by the Russian Foundation for Basic Research, grants No. 17-03-00111 and 16-53-52012. V.Yu.V. appreciates the support from the European Regional Development Fund, project TK134. D.K. acknowledges funding by the Deutsche Forschungsgemeinschaft (DFG) within Schwerpunktprogramm (SPP) 1386. A.A.T. is grateful for the financial support by the Federal Ministry for Education and Research under the Sofja Kovalevskaya Award of the Alexander von Humboldt Foundation. This work is partially based on the experiments performed at the Swiss spallation neutron source SINQ, Paul Scherrer Institute (PSI), Villigen, Switzerland. The authors thank Dr. Denis Sheptyakov and Dr. Matthias Frontzek for their help during the the NPD experiments.
\end{acknowledgements}

\bibliography{fulltext}
\end{document}